\documentclass[prl,aps,amsfonts,amssymb,amsmath,floats,floatfix,twocolumn,superscriptaddress,unsortedaddress]{revtex4}
\usepackage{epsfig,graphicx}
\usepackage{graphics}
\usepackage{float}
\begin{document}

\title{Surface-enhanced charge-density-wave instability in underdoped Bi2201}

\author{J.A. Rosen} \thanks{J.A.R. and R.C. coequally authored this work.}
\affiliation{Department of Physics {\rm {\&}} Astronomy, University of British Columbia, Vancouver, British Columbia V6T\,1Z1, Canada}
\author{R. Comin} \thanks{J.A.R. and R.C. coequally authored this work.}
\affiliation{Department of Physics {\rm {\&}} Astronomy, University of British Columbia, Vancouver, British Columbia V6T\,1Z1, Canada}
\author{G. Levy}
\affiliation{Department of Physics {\rm {\&}} Astronomy, University of British Columbia, Vancouver, British Columbia V6T\,1Z1, Canada}
\affiliation{Quantum Matter Institute, University of British Columbia, Vancouver, British Columbia V6T\,1Z4, Canada}
\author{D. Fournier}
\affiliation{Department of Physics {\rm {\&}} Astronomy, University of British Columbia, Vancouver, British Columbia V6T\,1Z1, Canada}
\author{Z.-H. Zhu}
\affiliation{Department of Physics {\rm {\&}} Astronomy, University of British Columbia, Vancouver, British Columbia V6T\,1Z1, Canada}
\author{B. Ludbrook}
\affiliation{Department of Physics {\rm {\&}} Astronomy, University of British Columbia, Vancouver, British Columbia V6T\,1Z1, Canada}
\author{C.N. Veenstra}
\affiliation{Department of Physics {\rm {\&}} Astronomy, University of British Columbia, Vancouver, British Columbia V6T\,1Z1, Canada}
\author{A. Nicolaou}
\affiliation{Department of Physics {\rm {\&}} Astronomy, University of British Columbia, Vancouver, British Columbia V6T\,1Z1, Canada}
\affiliation{Quantum Matter Institute, University of British Columbia, Vancouver, British Columbia V6T\,1Z4, Canada}
\author{\\D.\,Wong}
\affiliation{Department of Physics {\rm {\&}} Astronomy, University of British Columbia, Vancouver, British Columbia V6T\,1Z1, Canada}
\author{P. Dosanjh}
\affiliation{Department of Physics {\rm {\&}} Astronomy, University of British Columbia, Vancouver, British Columbia V6T\,1Z1, Canada}
\author{Y. Yoshida}
\affiliation{National Institute of Advanced Industrial Science and Technology (AIST), Tsukuba, 305-8568, Japan}
\author{H. Eisaki}
\affiliation{National Institute of Advanced Industrial Science and Technology (AIST), Tsukuba, 305-8568, Japan}
\author{G.R. Blake}
\affiliation{Materials\,Science\,Centre,\,University\,of\,Groningen,\,Nijenborgh\,4,\,9747\,AG\,Groningen,\,The\,Netherlands}
\author{F. White}
\affiliation{Agilent Technologies UK Ltd., 10 Mead Road, Oxford Industrial Park, Yarnton, Oxfordshire, OX5 1QU, UK}
\author{T.T.M. Palstra}
\affiliation{Materials\,Science\,Centre,\,University\,of\,Groningen,\,Nijenborgh\,4,\,9747\,AG\,Groningen,\,The\,Netherlands}
\author{R. Sutarto}
\affiliation{Canadian Light Source, University of Saskatchewan, Saskatoon, Saskatchewan S7N\,0X4, Canada}
\author{\\F. He}
\affiliation{Canadian Light Source, University of Saskatchewan, Saskatoon, Saskatchewan S7N\,0X4, Canada}
\author{A. Fra\~{n}o Pereira}
\affiliation{Max Planck Institute for Solid State Research, Heisenbergstrasse 1, D-70569 Stuttgart, Germany}
\affiliation{Helmholtz-Zentrum Berlin f\"{u}r Materialien und Energie,
Wilhelm-Conrad-R\"{o}ntgen-Campus BESSY II, Albert-Einstein-Str. 15, D-12489 Berlin, Germany}
\author{Y. Lu}
\affiliation{Max Planck Institute for Solid State Research, Heisenbergstrasse 1, D-70569 Stuttgart, Germany}
\author{B. Keimer}
\affiliation{Max Planck Institute for Solid State Research, Heisenbergstrasse 1, D-70569 Stuttgart, Germany}
\author{G. Sawatzky}
\affiliation{Department of Physics {\rm {\&}} Astronomy, University of British Columbia, Vancouver, British Columbia V6T\,1Z1, Canada}
\affiliation{Quantum Matter Institute, University of British Columbia, Vancouver, British Columbia V6T\,1Z4, Canada}
\author{L. Petaccia}
\affiliation{Elettra Sincrotrone Trieste, Strada Statale 14 Km 163.5, 34149 Trieste, Italy}
\author{A. Damascelli}
\affiliation{Department of Physics {\rm {\&}} Astronomy, University of British Columbia, Vancouver, British Columbia V6T\,1Z1, Canada}
\affiliation{Quantum Matter Institute, University of British Columbia, Vancouver, British Columbia V6T\,1Z4, Canada}

\maketitle

{\bf Neutron and X-ray scattering experiments have provided mounting evidence for spin and charge ordering phenomena in underdoped cuprates. These range from  early work on stripe correlations in Nd-LSCO to the latest discovery of charge-density-waves in YBCO. Both phenomena are characterized by a pronounced dependence on doping, temperature, and an externally applied magnetic field. Here we show that these electron-lattice instabilities exhibit also a previously unrecognized bulk-surface dichotomy. Surface-sensitive electronic and structural probes uncover a temperature-dependent evolution of the CuO$_2$ plane band dispersion and apparent Fermi pockets in underdoped Bi2201, which is directly associated with an hitherto-undetected strong temperature dependence of the incommensurate superstructure periodicity below 130\,K. In stark contrast, the structural modulation revealed by bulk-sensitive probes is temperature independent. These findings point to a surface-enhanced incipient charge-density-wave instability, driven by Fermi surface nesting. This discovery is of critical importance in the interpretation of single-particle spectroscopy data and establishes the surface of cuprates and other complex oxides as a rich playground for the study of electronically soft phases.} 
\begin{figure*}[t!]
\includegraphics[width=0.95\linewidth]{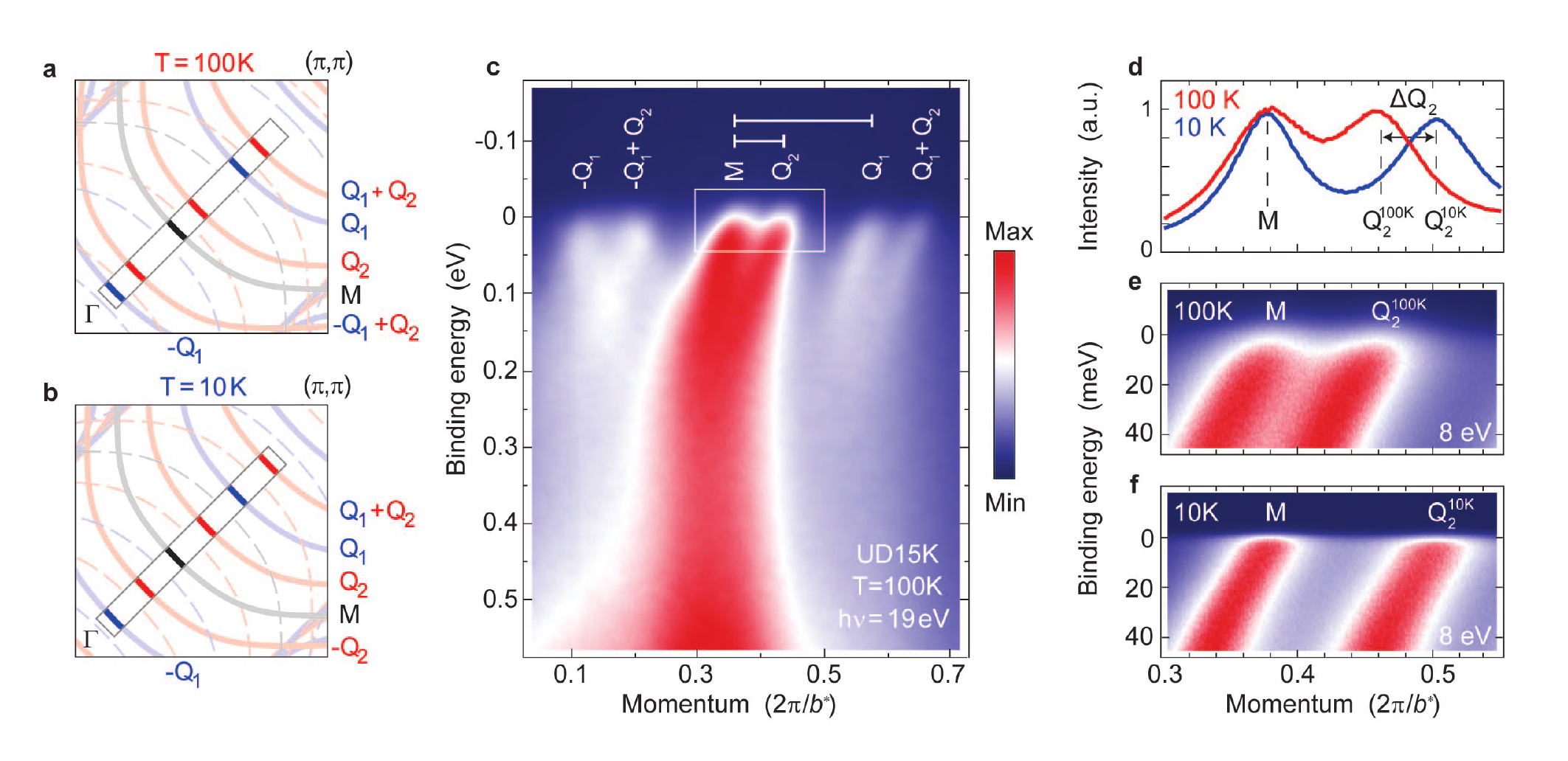}
\vspace{-0.2cm}
\caption{{\bf Temperature dependence of the nodal electronic structure of UD15K Bi2201}. (a,b) Sketch of one quadrant of the tetragonal Brillouin zone for $T\!=\!100$ and 10\,K, respectively; indicated are the expected Fermi surfaces belonging to the main band (M) and its replicas due to different {\bf Q}$_1$ and {\bf Q}$_2$ superstructure vector combinations (solid lines), as well as all the corresponding backfolded features due to the orthorhombicity of the crystal (dashed lines, so-called `shadow-bands'). The nodal strip in (a,b) highlights the region measured by ARPES with various photon energies and temperatures in (c,e,f), and the six bands detected for this experimental geometry and polarization (the photon polarization is set in the plane of detection to suppress all but the main band and its replicas). Momentum distribution curves (MDCs) at $E_{\mathrm{F}}$ for 10 and 100\,K are directly compared in (d). Also note that in (c-f), as throughout the paper, momentum axes are expressed in units of $2\pi/a^*$ and $2\pi/b^*$, where $a^*\!\simeq\!b^*\!\simeq\!\sqrt{2}\!\times\!3.86\mathrm{\AA}$ refer to the orthorhombic unit cell of Bi2201 (3.86\AA\ is the in-plane Cu-O-Cu distance).}
\label{ARPES_nodal}
\end{figure*}

The underdoped cuprates, with their pseudogap phenomenology \cite{timuskPG,normanPG,hufner} and marked departure from Fermi liquid behavior \cite{FournierNP}, have led to proposals of a wide variety of possible phases ranging from conventional charge and magnetic order to nematic and unconventional density wave instabilities \cite{zaanen1989,tranquada1995,varma1997,kivelson1998,sudip2001,kaminski2002,hoffman2002,howald2003,vershinin2004,shen2005,fauque2006,koshaka2007,doiron2007,xia2008,li2008,wise2008,hinkov2008,grilli2009,daou2010,lawler2010,parker2010,he2011,mesaros2011}. Despite the extensive theoretical and experimental effort, the generic phase behavior of the underdoped cuprates is still a matter of heated debate, primarily because of the lack of an order parameter that could be universally associated with the underdoped regime of the high-$T_{\mathrm{c}}$ cuprates (HTSCs). For instance, early on evidence was obtained for long-range spin and charge order in the form of uniaxial stripes \cite{tranquada1995}. This phenomenology has been detected in compounds belonging to the La$_{2-x-y}$(Sr,Ba)$_{x}$(Nd,Eu)$_{y}$CuO$_{4}$ family \cite{tranquada1995,fujita2004,abbamonte2005,birgeneau2006}, namely Eu-LSCO, Nd-LSCO, LBCO, and recently also pristine LSCO (where stripe order appears as a near-surface effect \cite{Wu2012}), and it is historically associated with the family-specific reduction of superconducting $T_{\mathrm{c}}$ near 12\% doping, the so-called `1/8-anomaly'.

More recently, high-field quantum oscillations \cite{doiron2007}, Hall resistance \cite{leboeuf2007} and thermoelectric transport \cite{laliberte2011} results on underdoped YBa$_2$Cu$_3$O$_{6+x}$ (YBCO) were interpreted as a signature of a magnetic-field induced reconstruction of the normal-state Fermi surface, suggesting that stripe order and/or a charge-density-wave (CDW) phase might be more general features of HTSCs' underdoped regime. Interest in this direction has been burgeoning with the latest nuclear-magnetic resonance (NMR) \cite{wu2011}, resonant X-ray scattering (REXS) and X-ray diffraction (XRD) results \cite{ghiringhelli2012,chang2012,achkar2012}, providing direct evidence for a long-range incommensurate CDW in YBCO around 10-12\% hole doping, which further shows a suppression for $T\!<\!T_{\mathrm{c}}$ and an enhancement with increasing magnetic field. Although this phenomenon bears some differences with respect to charge stripes, a common intriguing aspect is that they both are electronically-driven forms of ordering and appear to compete with superconductivity.

If a CDW phase in underdoped cuprates is universal, it should be observable in compounds with similar doping levels regardless of their structural details. In addition, it is of fundamental importance to connect structural observations (XRD, REXS) to those of electronic probes, such as angle-resolved photoemission (ARPES) and scanning tunnelling (STM) spectroscopy. However, for YBCO this might be prevented altogether by the polar instability and self-doping of the (001) surface; in fact, ARPES studies have not yet directly detected a folding of the electronic band structure \cite{FournierNP,Hossain} carrying the signature of a symmetry-broken CDW state as otherwise seen in either quantum oscillation \cite{doiron2007} or X-ray diffraction experiments \cite{ghiringhelli2012,chang2012,achkar2012}. To broaden the search and attempt this connection, the most interesting family is the one of Bi-cuprates which, owing to their extreme two-dimensionality and natural cleavage planes, have been extensively studied by single-particle spectroscopies \citep{damascelliRMP,fischerRMP}. ARPES and STM have provided rich insight into the electronic properties of the CuO$_2$ plane, including signatures of broken symmetries \cite{kaminski2002,vershinin2004,hanaguri2004,shen2005,koshaka2007,lawler2010,mesaros2011,he2011} and hints of a `pseudogap phase-transition' \cite{he2011}, although the identification of a \textit{bona fide} order parameter has remained elusive. More specifically in regards to a potentially underlying CDW instability, pristine Bi-cuprates have been shown to exhibit multiple superstructures, and while some of these modulations originate from the structural mismatch between BiO and CuO$_2$ lattice planes and hence are non-electronic in origin \cite{damascelliRMP,structure2,structure1,mans2006,phil}, others have been recognized by STM to evolve strongly with doping and magnetic field \cite{hoffman2002,howald2003,wise2008,parker2010}; however, their relationship to the `structural' superstructures and the Fermi surface has remained unclear. Our experimental results will provide new and surprising insight in this direction.

Here we study the structural and electronic properties of Bi$_2$Sr$_{2-x}$La$_x$CuO$_{6+\delta}$ (Bi2201), whose crystal structure exhibits a stacking of well-spaced, single CuO$_2$ layers in the unit cell and a highly-ordered superstructure \citep{phil}, by means of surface-sensitive photoemission spectroscopy (ARPES) and low-energy electron diffraction (LEED) probes, as well as bulk-sensitive resonant (REXS) and non-resonant (XRD) X-ray diffraction. We focus on the temperature dependence of the electronic structure from under ($p\!\simeq\!0.12$, $T_{\mathrm{c}}\!=\!15$\,K, UD15K) to nearly optimal doping ($p\!\simeq\!0.16$, $T_{\mathrm{c}}\!=\!30$\,K, OP30K). We discover a temperature dependent evolution of the CuO$_2$ plane band dispersion and apparent Fermi surface pockets, which is directly associated with the evolution of the incommensurate superstructure. Surprisingly, this effect is limited to the surface (ARPES-LEED), with no corresponding temperature evolution in the bulk (XRD-REXS). The quasilinear, continuous variation of the surface modulation wavelength $2\pi/Q_2$ from $\sim\!66$ to $43\mathrm{\AA}$, below a characteristic $T_{\mathrm{{\bf Q}}_2}\!\simeq\!130$\,K, provides evidence for a surface-enhanced CDW instability, driven by the interplay of nodal and antinodal Fermi surface nesting. 
\begin{figure*}[t!]
\includegraphics[width=1\linewidth]{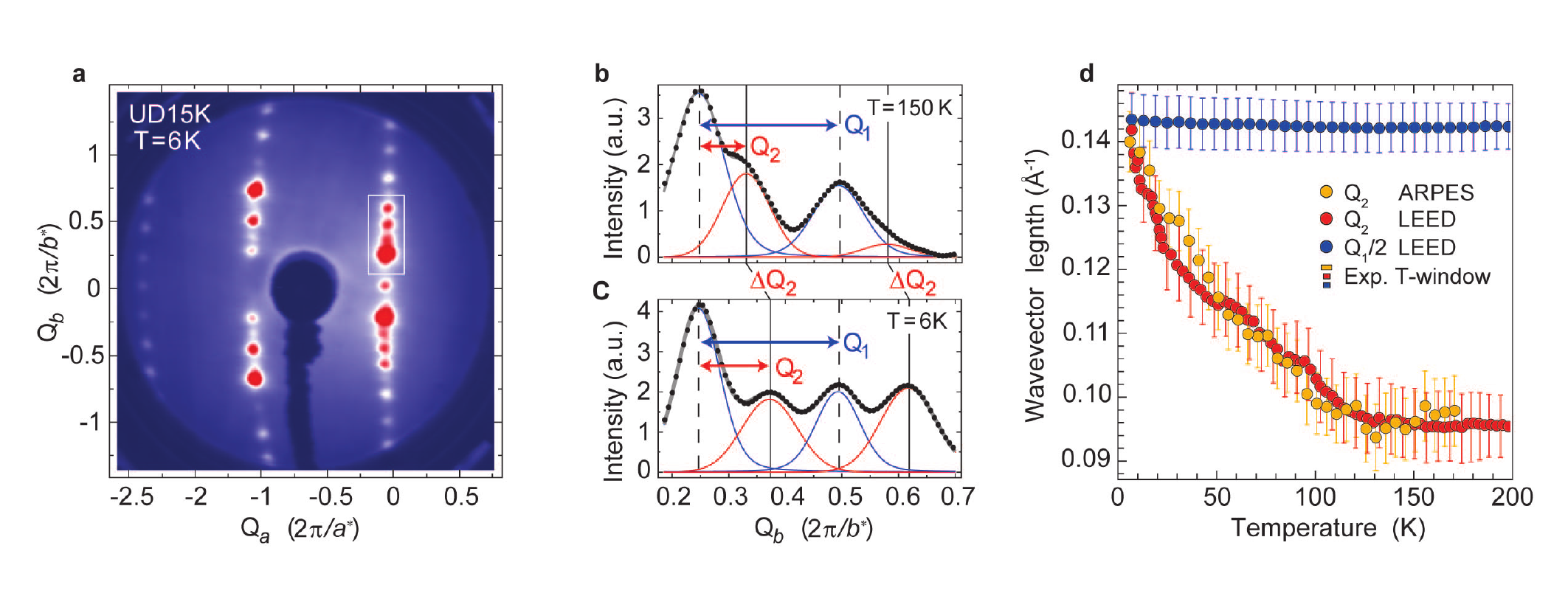}
\vspace{-1.2cm}
\caption{{\bf Temperature dependence of the superstructure modulations of UD15K Bi2201}. (a) Typical LEED pattern measured at $T\!=\!6$\,K. The rectangular box in (a) highlights the region shown in detail for $T\!=\!150$ and 6\,K in (b) and (c), respectively. In (b,c) symbols represent the data from a vertical cut along the center of the box in (a), while blue and red curves are a Voigt fit of the {\bf Q}$_1$ and {\bf Q}$_2$ superstructure peaks. (d) Magnitude of the {\bf Q}$_1$ and {\bf Q}$_2$ superstructure vectors in \AA$^{-1}$ versus temperature, as inferred from LEED and ARPES-MDC analysis at 21\,eV photon energy (and in agreement with ARPES from 7 to 41\,eV, see e.g. Fig.\,\ref{ARPES_nodal}). The yellow and red/blue boxes indicate the temperature integration window of each data point, for ARPES (5\,K) and LEED (3\,K), respectively, and the error bars show the goodness of fit for the Voigt profiles determined from a Chi-squared test. Note that, for the almost temperature independent {\bf Q}$_1$, half of the actual value is plotted for a more direct comparison with {\bf Q}$_2$ and only the LEED data are shown (the ARPES data are equivalent and thus omitted).}
\label{LEED_T}
\end{figure*}

\section{Results}

\noindent {\bf Orthorhombic and modulated structure of the Bi-cuprates.} 
An important aspect to consider for the study of Bi-cuprates is that these materials are not structurally tetragonal, but instead orthorhombic, with 2 inequivalent Cu atoms per CuO$_2$ plane \cite{damascelliRMP,structure2,structure1,mans2006,phil}. This leads to a 45$^\circ$\,degree rotated and $\sqrt{2}\!\times\!\sqrt{2}\!\times\!1$ larger unit cell, as compared to the tetragonal one, with lattice parameters $a^*\!\cong\!b^*\!\cong\!\sqrt{2}\!\times\!\mathrm{3.86}\mathrm{\AA}$, where 3.86\AA\ is the planar Cu-O-Cu distance ($c\!\cong\!\mathrm{24.9}\mathrm{\AA}$ for both structures). Note that throughout the paper we refer to the orthorhombic unit cell, with momentum axes expressed using the reciprocal lattice units (r.l.u.) $2\pi/a^*$, $2\pi/b^*$ and $2\pi/c$. The orthorhombicity and consequent band backfolding have been shown to be responsible for the observation of the so-called ``shadow bands'' \cite{mans2006}, a replica of the hole-like CuO$_2$ Fermi surface centered at the $\Gamma$ point, thus settling a longstanding debate on their possible antiferromagnetic origin \cite{aebi1994}. In addition, the presence of  incommensurate superstructure modulations, arising from a slight lattice mismatch between the BiO layers and the CuO$_2$ perovskite blocks \cite{Zhiqiang1997}, further adds to the complexity of the Fermi surface. As for single-layer Bi2201 specifically, while a single {\bf Q}$_1$ superstructure vector is known to give rise to additional folded replicas along the orthorhombic $b^{*}$ axis at optimal doping (OP) \cite{damascelliRMP}, two distinct structural modulations with {\bf Q}$_1$ and {\bf Q}$_2$ wavevectors arise with underdoping (UD). If these complications are not fully taken into account in analyzing ARPES data, the resulting highly complex Fermi surface appears to be composed of a small set of closed pockets \cite{phil}.\\
\begin{figure*}[t!]
\includegraphics[width=0.95\linewidth]{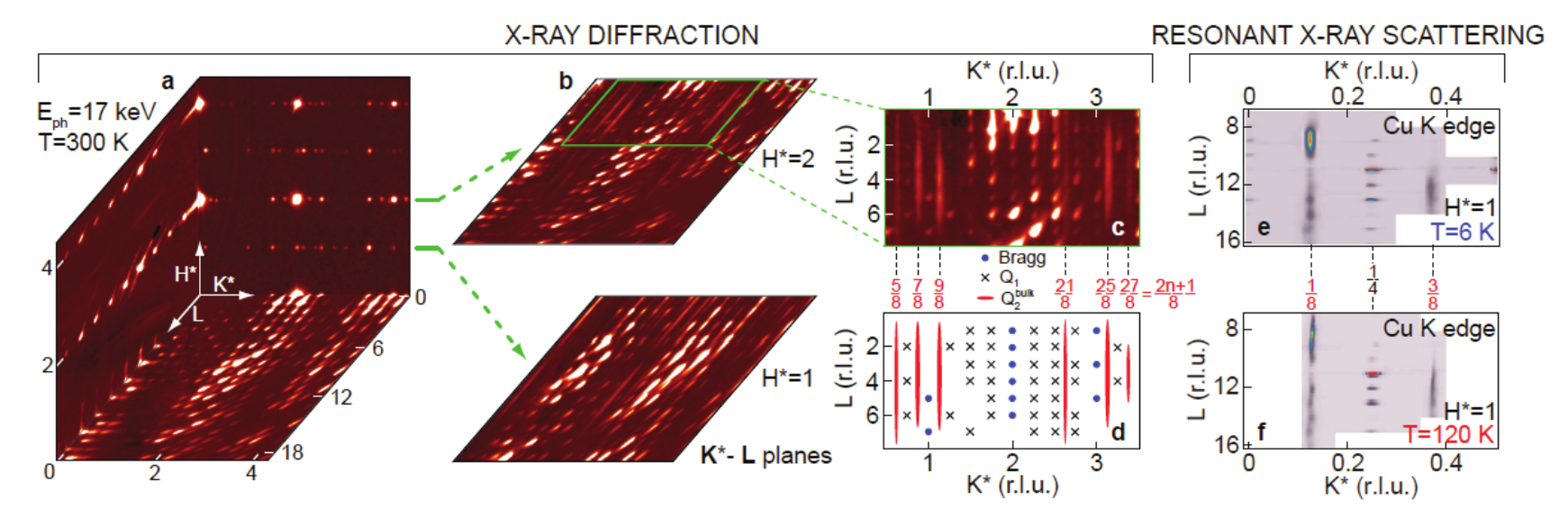}
\vspace{-0.2cm}
\caption{{\bf X-ray measurements of the superstructure modulations in UD15K Bi2201}. (a) 3D view of the basal planar sections of XRD maps at ${E}_{\mathrm{ph}} \sim 17$\,keV (only positive axes are shown). (b) ${H}^{*}\!\!=\,$1,\,2 slices, showing the appearance of period-8 diffraction rods, while no similar features are found for ${H}^{*}\!\!=\,$0 [see bottom plane in panel (a)], thus explaining the lack of period-8 features in soft X-ray REXS. (c) Enlarged view of corresponding region of interest in (b) for the $H{}^{*}\!\!=\,$2 slice. (d) Schematic cartoon explaining the multiple features that are visible in (c): blue circles correspond to Bragg peaks for integer ${K}^{*}$ and \textit{L} orders; black crosses are the 1/4-order {\bf Q}$_{1}$ peaks; red ellipses are the 1/8-order {\bf Q}$_{2}$  peaks. (e) REXS map acquired at the Cu-\textit{K} edge (${E}_{\mathrm{ph}} \sim 8.9$ keV) at 6 K, representing a {\bf K}$^{*}$-{\bf L} plane at ${H}^{*}\!\!=\,$1. (f) Same as (e), but acquired at 120 K.}
\label{xray}
\end{figure*}

\noindent {\bf Probing the surface with ARPES and LEED.} \linebreak
We begin with the discussion of the UD15K Bi2201 ARPES data from along the nodal direction presented in Fig.\,\ref{ARPES_nodal}. As demonstrated in previous work  \cite{phil}, and here sketched in Fig.\,\ref{ARPES_nodal}(a,b) for a simpler identification of the various bands, the high crystallinity of these samples allows resolving the Fermi surface of Bi-cuprates to an unprecedented level of detail: the main (M) CuO$_2$-plane band (black solid line), its {\bf Q}$_1$ and {\bf Q}$_2$ superstructure replicas stemming from the BiO-layer-induced incommensurate superstructure (red and blue solid lines), and all the corresponding backfolded bands  due to the orthorhombicity of the crystal (dashed lines). 
Furthermore, as shown in Fig.\,\ref{ARPES_nodal}(c) for UD15K at $T\!=\!100$\,K, and emphasized in the highlighted nodal strip in Fig.\,\ref{ARPES_nodal}(a), by taking advantage of the polarization-dependent selection rules \cite{phil} one can selectively suppress the redundant backfolded bands to highlight more cleanly the behavior of main (M) and {\bf Q}$_1$-{\bf Q}$_2$ bands. The ability to simultaneously detect all superstructure replicas allows us to uncover -- in the temperature dependence -- a new and unprecedented aspect of the data: while the position of the main CuO$_2$ band is completely temperature independent, between 100 and 10\,K there is a significant shift in momentum of only (and all) the {\bf Q}$_2$-related bands [see Fig.\,\ref{ARPES_nodal}(e) and (f), and Fig.\,\ref{ARPES_nodal}(d) for the direct comparison between 10-100\,K momentum distribution curves (MDCs) at $E_{\mathrm{F}}$]. This is summarized in the 10\,K Fermi surface sketch of Fig.\,\ref{ARPES_nodal}(b), which illustrates that a critical consequence of this effect is a seeming volume change of all ostensible Fermi surface pockets defined by the various backfolded bands, despite the fact that the actual number of carriers is not changing at all.

The ARPES results are complemented by a detailed analysis of the superstructure diffraction vectors from LEED. On UD15K at 6\,K, rather than individual Bragg peaks [Fig.\,\ref{LEED_T}(a)], the experiment gives  lines of {\bf Q}$_1$ and {\bf Q}$_2$ fractional spots along the orthorhombic $b^*$ axis. From the fit of the LEED data [Fig.\,\ref{LEED_T}(c)], we obtain for the magnitude of the superstructure wavevectors the values $Q_1^{\mathrm{6K}}\!=\!0.285\!\pm\!0.015\textrm{\AA}^{-1}$ and $Q_2^{\mathrm{6K}}\!=\!0.142\!\pm\!0.015\textrm{\AA}^{-1}$, corresponding to $\sim\!1/4$ and 1/8 in r.l.u., respectively. Also LEED, on this highly-resolved superstructure, reveals a remarkable temperature dependence [Fig.\,\ref{LEED_T}(b,c)]. Consistent between LEED and ARPES-MDC analysis [Fig.\,\ref{LEED_T}(d)], while $Q_1$ is virtually temperature independent from 5 to 300\,K, $Q_2$ increases with respect to its high-temperature value $Q_2^{\mathrm{300K}}\!=\!0.095\!\pm\!0.015\textrm{\AA}^{-1}$ ($\sim\!1/12$ in r.l.u.) below a $T_{\mathrm{Q}_2}\!\simeq\!130$\,K. The evolution of $Q_2$ -- as seen by both electronic and structural probes -- implies an inter-unit-cell structural and/or electronic modulation, with a wavelength $2\pi/Q_2$ evolving from 66 to 44\,\AA\ (i.e. from $~\!12$ to $8\times b^*$) upon cooling from 130 down to 5\,K. \\

\noindent {\bf Bulk-sensitivity with XRD and REXS.}
The surface sensitivity of ARPES and LEED calls for an investigation of the same phenomenology by means of light scattering techniques, which are known to probe materials deeper in the bulk. In the following discussion, we will refer to reciprocal space coordinates as {\bf H}$^{*}$, {\bf K}$^{*}$, {\bf L} (representing the reciprocal axes of respectively {\bf a}$^{*}$, {\bf b}$^{*}$, {\bf c}), and reciprocal lattice units will be used. At all photon energies it is possible to clearly identify the supermodulation associated with {\bf Q}$_1$. In particular, REXS maps taken on UD15K Bi2201 at the Cu, La, and O soft X-ray edges all exhibit a clear enhancement at this wavevector (see Supplementary Note 1 for details). This confirms that the corresponding modulation is present throughout the unit cell, and therefore also in the CuO${}_{2}$ plane, explaining the strong folded replicas observed in ARPES. However, modulations with longer periods are not detected. These can be probed by XRD maps measured at 17\,keV photon energy, thus revealing a much larger portion of reciprocal space, as shown in Fig.\,\ref{xray}(a-c) for $T\!=\!300$\,K. The {\bf H}$^{*}$-{\bf K}$^{*}$ section in Fig.\,\ref{xray}(a), which can be compared to the LEED map in Fig.\,\ref{LEED_T}(a), also features a multitude of superstructure satellite peaks along {\bf K}$^{*}$ (the Bragg peaks being the most intense ones). Fig.\,\ref{xray}(b) displays the {\bf K}$^{*}$-{\bf L} sections for $H^{*}\!=\!1,2$, which reveal the presence of new features exhibiting a peculiar elongation along {\bf L} and  period-8 modulation along {\bf K}$^{*}$ with positions given by ${K}^{*}\!=\!(2n+1)/8$. The latter are therefore incompatible with the near period-4  modulation associated with {\bf Q}$_{1}$ or any of its harmonics (also note that no similar features are found for ${H}^{*}\!=\!0$, thus explaining the lack of period-8 rods in soft X-ray REXS, which due to kinematic constraints can only probe a reduced portion of reciprocal space). Different orders of this period-8 modulation can be seen when zooming in to the ${H}^{*}\!=\!2$ slice in Fig.\,\ref{xray}(c), with their assignment given more schematically in Fig.\,\ref{xray}(d). These are located at positions ${\mathbf{Q}}_{2}^{ij}\!=\!n\,\mathbf{G} \pm i\,{\mathbf{Q}}_{1} \pm j\,{\mathbf{Q}}_{2}$, where $ \mathbf{G} $ is a reciprocal lattice vector, $ {\mathbf{Q}}_{1}\!=\!1/4\,{\hat{\mathrm{\bf u}}}_{{\mathrm{K}}^{*}}$, and $ {\mathbf{Q}}_{2}\!=\!1/8\,{\hat{\mathrm{{\bf u}}}}_{{\mathrm{K}}^{*}}$ (corresponding to the ARPES and LEED low-temperature {\bf Q}$_2^{\mathrm{6K}}$ value). Notably, the same features can be seen in resonant scattering at Cu and Bi deeper edges (i.e., in the hard X-ray regime). Fig.\,\ref{xray}(e,f) show corresponding {\bf K}$^{*}$-{\bf L} sections (${H}^{*}\!=\!1$), taken at the Cu-\textit{K} edge at low (6\,K) and high (120\,K) temperature, respectively. Additional data for the Bi-${L}_{3}$ edge and the temperature-dependent XRD maps are shown in Supplementary Note 1.

The intensity of the ${\mathbf{Q}}_{2}^{ij}$ rods is approximately 1 order of magnitude smaller than the most intense {\bf Q}$_1$ peak, within the same {\bf K}$^{*}$-{\bf L} sections. Considering the large probing depth of hard X-rays, this intensity ratio is too large to identify these as crystal truncation rods, or ascribe them to surface modulations. These period-8 spots therefore originate from an additional supermodulation which must be present in the bulk of the material, and characterized by poor c-axis coherence, as the elongated structure suggests. On the other hand, these features exhibit long-range order in the {\bf a}$^{*}$-{\bf b}$^{*}$ plane, as evidenced by their well-defined shape in {\bf H}$^{*}$-{\bf K}$^{*}$ sections, with correlation lengths $ \xi > 100\times b^{*} $. 

To summarize the findings from XRD and REXS on UD15K, no significant temperature dependence is observed between 300 and 6\,K in all scans, neither in the peak positions nor in the relative intensities. Altogether, these results suggest a scenario involving the presence of an additional bulk supermodulation with a well-defined periodicity along {\bf b}$^{*}$ ($\sim 8$ lattice periods), stable over a broad range of temperatures, and characterized by large correlation lengths within the (001) planes but poor coherence perpendicular to them. 

\section{Discussion}

The combination of surface (ARPES, LEED) and bulk (XRD, REXS) sensitive probes has enabled us to establish that {\bf Q}$_1$ and {\bf Q}$_2$ superstructure modulations are present both in the bulk and at the surface of underdoped Bi2201, close to 1/8 doping (UD15K). In addition, we have uncovered an unprecedented bulk-surface dichotomy in the temperature dependence of the superstructure modulations and corresponding electronic structure. While no dependence is observed for the {\bf Q}$_1$ and {\bf Q}$_2$ superstructure in the bulk and also for {\bf Q}$_1$ at the surface, we detected a pronounced temperature evolution associated with the surface $\mathrm{{\bf Q}}_2^{\mathrm{surf}}$. As for the doping dependence of this phenomenon, while the {\bf Q}$_1$ modulation survives all the way to optimal doping ($Q_1\!\simeq\!0.280$ and $0.273\mathrm{\AA}^{-1}$ for UD23K and OP30K, respectively), the {\bf Q}$_2$ modulation is substantially weakened and temperature independent for  UD23K ($p\!=\!0.14$, $Q_2\!\simeq\!0.135\mathrm{\AA}^{-1}$), and can no longer be detected in either LEED or ARPES on OP30K ($p\!>\!0.16$). This is discussed in the Supplementary Note 1 based on the doping and temperature dependent LEED, ARPES, and X-ray data.

\begin{figure*}[t!]
\includegraphics[width=1\linewidth]{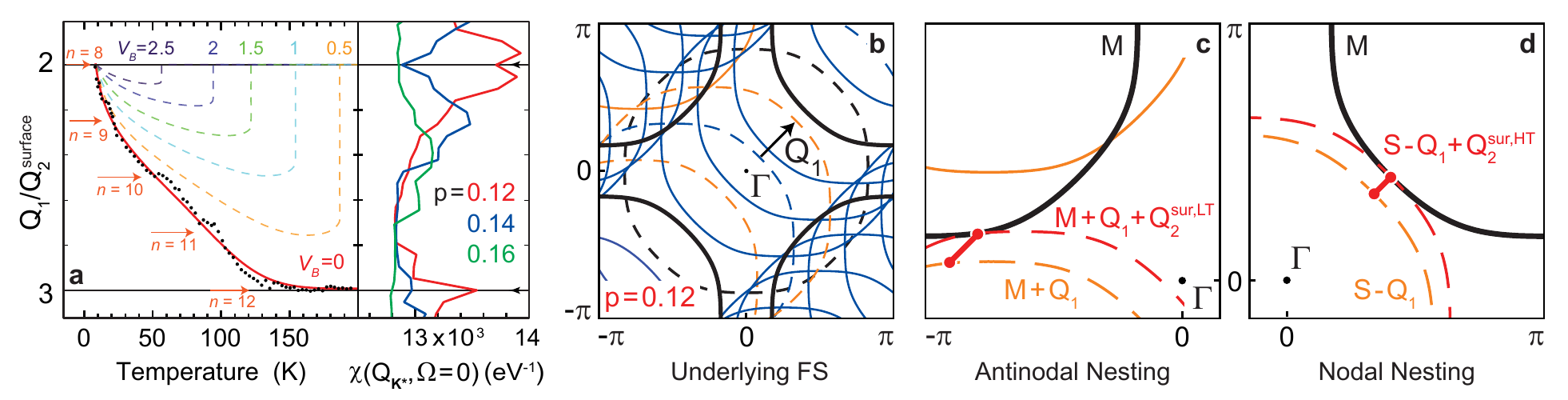}
\vspace{-0.2cm}
\caption{{\bf $\mathbf{\mathrm{{\bf Q}}_2}$ mean-field theory and nesting effects in the electronic susceptibility}. (a, main panel) Temperature evolution of the $Q_1/Q_2^{\mathrm{surf}}$ wavevector magnitude ratio (black dots), as inferred from the LEED data in Fig.\,\ref{LEED_T}, compared to the evolution of the mean-field predicted wavevector that minimizes the free energy (Supplementary Note 2); red arrows mark those wavevectors at which the modulation associated with {\bf Q}$_2$ becomes commensurate with the underlying orthorhombic lattice $Q_2\!=\!(2\pi/b^*)/n$, for various values of $n$. The colored curves illustrate the effect of increasing $V_\mathrm{B}$, showing how the bulk structure can pin the CDW and suppress the temperature dependence of {\bf Q}$_2$. (a, side panel) Calculated electronic susceptibility $ \chi (\mathbf{Q}) $, cut along the {\bf K}$^{*}$ direction in reciprocal space, for $ p\!=\! 0.12 $ (red), 0.14 (blue) and 0.16 (green). (b) Cartoon of the Fermi surface modeling used in the calculation of $ \chi (\mathbf{Q}) $; orange traces mark the Fermi surfaces involved in the nesting mechanism ($\mathrm{M+{Q}_{1}}$ and $\mathrm{S-{Q}_{1}}$). (c,d) Schematics of the antinodal and nodal nesting instabilities, which connect the main (M, black trace) band with $\mathrm{M + {Q}_{1}}$ and $\mathrm{S - {Q}_{1}}$ (orange traces), respectively. The resulting nesting vectors {\bf Q}$_{2}^{\mathrm{surf},\mathrm{LT}}$ and {\bf Q}$_{2}^{\mathrm{surf},\mathrm{HT}}$ are represented by the thick red connectors, while the corresponding {\bf Q}$_2$-derived Fermi surfaces are shown in dashed red.}
\label{ARPES_summary}
\end{figure*}

The dependence of the $Q_1/Q_2^{\mathrm{surf}}$ ratio versus temperature for UD15K is summarized in Fig.\,\ref{ARPES_summary}(a) and allows some important phenomenological observations: (i) the temperature dependence of $\mathrm{{\bf Q}}_2^{\mathrm{surf}}$ below $T_{\mathrm{{\bf Q}}_2}$ shows commensurability with the static {\bf Q}$_1$ modulation, as evidenced by the $Q_1/Q_2^{\mathrm{surf}}$ ratio varying from 3 to 2 over a range of 130\,K. (ii) The evolution of $Q_1/Q_2^{\mathrm{surf}}$ exhibits a possible transient lock-in behaviour when the wavelength of the $\mathrm{{\bf Q}}_2^{\mathrm{surf}}$ modulation is commensurate with the orthorhombic lattice:  $2\pi/Q_2^{\mathrm{surf}}\!=\!n \times b^*$, with $n$ ranging from 12 to 8, as marked by red arrows in Fig.\,\ref{ARPES_summary}(a) (see also Supplementary Note 2 for a more extended discussion). A similar albeit more pronounced behaviour has been observed for charge-stripe order in La$_2$NiO$_{4+\delta}$ from neutron scattering \cite{wochner1998}. (iii) In analogy to what was reported for manganites \cite{milward}, the continuous evolution of incommensurate wavevectors over a wide temperature range hints at competing instabilities, which can lead to a soft electronic phase. (iv) Finally, the fact that at low-temperature (LT) also $\mathrm{{\bf Q}}_2^{\mathrm{surf},\mathrm{LT}}\!\simeq\!\mathrm{{\bf Q}}_2^{\mathrm{bulk}}$ indicates a direct connection between the bulk and surface modulations.

We have succeeded to reproduce the details of the observed CDW instability and its temperature evolution using a twofold analysis (detailed in Supplementary Notes 2 and 3) involving: (i) the evaluation of the electronic susceptibility [through the zero-temperature, zero-frequency Lindhard function $ \chi (\mathbf{Q}, \Omega\!\!=\!\!0) $]; and (ii) a mean-field Ginzburg-Landau model based on an \textit{ad-hoc} phenomenological free-energy functional $ F\!\left[ \rho \right] $,  typical of that applied to CDW systems. The electronic susceptibility $ \chi (\mathbf{Q}, \Omega\!\!=\!\!0) $ has been calculated for various doping levels starting from an electronic structure comprised of main, shadow, and {\bf Q}$_1$-folded bands [see Fig.\,\ref{ARPES_summary}(b)] and is shown for $p\!=\!0.12$, 0.14, and 0.16 in the right-hand side panel of Fig.\,\ref{ARPES_summary}(a). Two peaks occur in the susceptibility along the {\bf K}$^{*}$ direction in reciprocal space at $Q_{{\mathrm{K}}^{*}}\!=\!0.095$ and 0.140\,$\mathrm{\AA}^{-1}$ for $p\!=\!0.12$, closely matching the {\bf Q}$_2$ supermodulation vectors for UD15K. This allows associating $\mathrm{{\bf Q}}_2^{\mathrm{surf},\mathrm{HT}}\!\simeq\!\mathrm{{\bf Q}}_1/3$ and $\mathrm{{\bf Q}}_2^{\mathrm{surf},\mathrm{LT}}\!\simeq\!\mathrm{{\bf Q}}_1/2$ with nodal and antinodal Fermi surface nesting, i.e. ${Q}^{\mathrm{surf},\mathrm{HT}}_{2}\!=\!{Q}^{\mathrm{N}}_{{\mathrm{K}}^{*}}\!=\!0.095$ and ${Q}^{\mathrm{surf},\mathrm{LT}}_{2}\!=\!{Q}^{\mathrm{AN}}_{{\mathrm{K}}^{*}}\!=\!0.140 \mathrm{\AA}^{-1}$ [these denominations designate the region in k-space where bands overlap maximally, as pictorially shown in Fig.\,\ref{ARPES_summary}(c,d)]. These nesting instabilities are very sensitive to the hole doping, especially for the steeper nodal dispersion; for $p\!=\!0.14$ and 0.16 the nodal peak abruptly vanishes, while the antinodal is split -- and therefore ceases to be commensurate to {\bf Q}$_1$ -- and gradually reduced and broadened towards optimal doping, yielding a progressively less pronounced instability. These findings qualitatively explain the experimentally-observed progressive weakening of the features associated with {\bf Q}$_2$ as hole doping is increased (discussed in Supplementary Note 1).
Ultimately, this establishes the specific high- and low-temperature values observed for the surface CDW modulation on UD15K, $\mathrm{{\bf Q}}_2^{\mathrm{surf},\mathrm{HT}}$ and $\mathrm{{\bf Q}}_2^{\mathrm{surf},\mathrm{LT}}$, to be associated with competing Fermi surface nesting instabilities of the {\bf Q}$_1$-modulated orthorhombic crystal structure. Most importantly, this identifies the temperature dependent {\bf Q}$_2$ surface CDW as a phenomenon limited to the underdoped regime, near 1/8 doping, consistent with our experimental observations.

As for the origin of the observed temperature dependence, the evolution of $Q_1/Q_2^{\mathrm{surf}}$ (ratio of wavevector magnitude) is well captured by a phenomenological Ginzburg-Landau description based on the minimization of the surface free energy functional $ F\!\left[ \rho \right] $, and is thus consistent with an incipient CDW instability at the surface. This is shown by the comparison of LEED and theoretical results (red trace) for the evolution of $Q_1/Q_2^{\mathrm{surf}}$ in UD15K, shown in Fig.\,\ref{ARPES_summary}(a). Commensurability to the susceptibility peaks ($Q_1/Q_2^{\mathrm{surf}}\!\!=\,$2 and 3) underpins the low- and high-temperature limits, while the free energy $ F\!\left[ \rho \right] $, in absence of a bulk potential, provides a modeling of the surface, and accounts for the temperature dependence of the {\bf Q}$_2$ wavevector. In Ginzburg-Landau mean-field theory, this can be understood as a consequence of the temperature dependent harmonic content of a non-sinusoidal CDW (see Supplementary Note 2 and Supplementary Discussion for more on this point), which here coincides with the $Q_1/Q_2$ commensurability effects.

In our Ginzburg-Landau description we can also include the effect of the bulk potential $V_B$ associated with the `static' $\mathrm{{\bf Q}}_2^{\mathrm{bulk}}$ modulation as determined by X-ray diffraction ($V_{\mathrm{B}}\!=\!|V_{\mathrm{Q}_2}|$, with $V_{\mathrm{Q}_2}$ as defined in Supplementary Note 2). As shown by the simulated colored traces in Fig.\,\ref{ARPES_summary}(a), incorporating this potential progressively causes the CDW to lock in to the bulk structural modulation wavevector $\mathrm{{\bf Q}}_2^{\mathrm{bulk}}\!=\!\mathrm{{\bf Q}}_1/2$ at ${V}_{\mathrm{B}} \sim 3.0$, thus suppressing the temperature-dependence. The two regimes $V_{\mathrm{B}}\!=\!0$ and $V_{\mathrm{B}}\!>\!3$ represent the temperature-dependent-surface and temperature-independent-bulk limiting cases, providing agreement with the results of ARPES-LEED on the surface and XRD-REXS for the bulk. Intermediate values of $V_{\mathrm{B}}$ describe the subsurface region, which shows a CDW with reduced dependence on temperature, and instability towards first-order lock-in transitions to the $\mathrm{{\bf Q}}_2^{\mathrm{bulk}}$ wavevector [see dashed traces in Fig.\,\ref{ARPES_summary}(a)].

In conclusion, the temperature dependent evolution of the CuO$_2$ plane band dispersion and {\bf Q}$_2$ superstructure on the highly-ordered Bi2201 surface can be understood to arise from the competition between nodal and antinodal Fermi surface nesting instabilities, which give rise to a dynamic, continuously evolving wavevector. This also indicates that such a remarkable electron-lattice coupling is directly related to the ordinary, static {\bf Q}$_1$ superstructure -- as a necessary precursor to Fermi surface nesting at the low- and high-temperature $\mathrm{{\bf Q}}_2^{\mathrm{surf}}$ -- and giving rise to commensurability effects. Since the nodal nesting-response is very sensitive to the hole-doping, this also explains why the surface temperature-dependence disappears towards optimal doping. This establishes the importance of surface-enhanced CDW nesting instabilities in underdoped Bi-cuprates, and reveals a so-far undetected bulk-surface dichotomy. The latter is responsible for many important implications, such as the temperature-dependent volume change of all apparent Fermi surface pockets in ARPES, and could play a hidden role in other temperature-dependent studies.

\section{Methods}

\noindent {\bf Sample preparation.} For this study we used two underdoped ($x\!=\!0.8$, $p\!\simeq\!0.12$, UD15K and $x\!=\!0.6$, $p\!\simeq\!0.14$, UD23K) and one optimally doped ($x\!=\!0.5$, $p\!\simeq\!0.16$, OP30K) Bi$_2$Sr$_{2-x}$La$_x$CuO$_{6+\delta}$ single crystals ($p$ is the hole doping per planar copper away from half-filling). The superconducting $T_{\mathrm{c}}\!=\!15$, 23, and 30\,K, respectively, were determined from in-plane resistivity and magnetic susceptibility measurements.  For UD15K we found $T^*\!\simeq\!190$\,K, based on the onset of the deviation of the resistivity-versus-temperature curve from the purely linear behavior observed at high temperatures. 
\\

\noindent {\bf ARPES and LEED experiments.} ARPES measurements were performed at UBC with 21.2\,eV photon energy (HeI), and at the Elettra synchrotron BaDElPh  beamline with photon energy ranging from 7 to 41\,eV. In both cases the photons were linearly polarized and the polarization direction -- horizontal ($p$) or vertical ($s$) -- could be varied with respect to the electron emission plane. Both ARPES spectrometers are equipped with a SPECS Phoibos 150 hemispherical analyzer; energy and angular resolution were set to 6-10\,meV and 0.1$^{\circ}$. The samples were aligned by conventional Laue diffraction prior to the experiments and then mounted with the in-plane Cu-O bonds either parallel or at 45$^{\circ}$ with respect to the electron emission plane. LEED measurements were performed at UBC with a SPECS ErLEED 100;  momentum resolution was set to 0.01\,\AA$^{-1}$ by using a low electron energy of 37\,eV, at which value the signal intensity reaches a maximum. During the LEED measurements, the samples were oriented with the orthorhombic $b^*$-axis vertical in reference to the camera, and rotated by $7\,^{\circ}$ in the horizontal plane to detect more spots. For both LEED and ARPES, the samples were cleaved \textit{in situ} at pressures better than 5$\times$10$^{-11}$\,torr. The detailed temperature-dependent experiments were performed on the UBC ARPES spectrometer, which is equipped with a 5-axis helium-flow cryogenic manipulator operating between 2.7 and 300\,K. The ARPES (LEED) data were acquired at 0.5\,frame/sec (30\,frame/sec), while the sample was cooled at a continuous rate of 0.1\,K/min (1\,K/min). The ARPES (LEED) data were averaged over 1,500 (5,400) images, resulting in ARPES spectra (LEED curves) with a temperature precision of 5\,K (3\,K). The higher temperature accuracy achieved in LEED stems from its tenfold signal-to-noise ratio as compared to ARPES. 
\\

\noindent {\bf Light scattering experiments.} Resonant elastic soft X-ray measurements were taken using a 4-circle diffractometer at the REIXS beamline  at the Canadian Light Source, working at the O-\textit{K} (${E}_{\mathrm{ph}} \sim 530$\,eV), La-${M}_{4,5}$ (${E}_{\mathrm{ph}} \sim 836$\,eV) and Cu-${L}_{2,3}$ (${E}_{\mathrm{ph}} \sim 930$\,eV) absorption edges. Hard X-ray scans were performed using a psi-8 diffractometer (8-circle) at the Mag-S  beamline at BESSY, working at the Cu-\textit{K} (${E}_{\mathrm{ph}} \sim 8.9$\,keV) and Bi-${L}_{3}$ (${E}_{\mathrm{ph}} \sim 13.2$\,keV) deep edges. Both soft and hard X-ray scattering measurements were performed in the temperature range 15-300\,K. X-ray diffraction reciprocal space maps were acquired using an Agilent Technologies SuperNova A diffractometer. The data were collected at 300\,K and 100\,K using Mo-${K}_{\mathrm{\alpha}}$  and  Cu-${K}_{\mathrm{\alpha}}$  radiation, respectively. The excitation energies used for these experiments correspond in turn to approximate attenuation lengths $ \alpha $ of $ \sim 150$\,nm (Cu-${L}_{2,3}$), 6\,$\rm{\mu}$m (Cu-\textit{K}) and 12\,$\rm{\mu}$m (Mo-${K}_{\mathrm{\alpha}}$). In all cases samples were pre-oriented using Laue diffraction and mounted {\bf b}$^{*}$ and {\bf c} axes in the scattering plane. In order to expose an atomically flat (001) surface, \textit{in}- and \textit{ex- situ} cleaving procedures were adopted for soft and hard X-ray measurements, respectively.

\providecommand{\noopsort}[1]{}\providecommand{\singleletter}[1]{#1}%

\subsection{Acknowledgments}

We gratefully acknowledge I.S. Elfimov, S.A. Kivelson, V. Hinkov, R.S. Markiewicz, M.R. Norman, L. Taillefer, and J.M. Tranquada for discussions, and D. Lonza and E. Nicolini for technical assistance at Elettra. This work was supported by the Max Planck -- UBC Centre for Quantum Materials, the Killam, Alfred P. Sloan, Alexander von Humboldt, and NSERC's Steacie Memorial Fellowship Programs (A.D.), the Canada Research Chairs Program (A.D. and G.A.S.), NSERC, CFI, and CIFAR Quantum Materials. The research at CLS is supported by NSERC, NRC, CIHR, and the University of Saskatchewan.

\subsection{Author Contributions}

J.A.R., R.C. and A.D. conceived this investigation -- J.A.R. carried out the temperature-dependence ARPES and LEED experiments at UBC, with assistance from G.L., B.L., Z.-H.Z. C.N.V., D.W., P.D. -- J.A.R. performed additional temperature-dependent ARPES at Elettra with assistance from R.C., D.F., L.P. -- R.C. performed REXS measurements at CLS with the assistance of R.S. and F.H., and at BESSY with A.F.P. and Y.L. -- G.R.B., F.W., and T.M.M.P. are responsible for the XRD measurements -- J.A.R., R.C., G.L., B.K., G.A.S., and A.D. are responsible for data analysis and interpretation -- J.A.R. is responsible for Ginzburg-Landau description -- Y.Y and H.E. grew and characterized the samples -- All of the authors discussed the underlying physics and contributed to the manuscript. A.D. is responsible for overall project direction, planning, and management.

\subsection{Author Information}

The authors declare no competing financial interests. Correspondence and requests for materials should be
addressed to A.D. (damascelli@physics.ubc.ca).

\end{document}